**Classification:** Biological sciences - Neuroscience

**Title:** Learning to make external sensory stimulus predictions using internal correlations in populations of neurons

***Short title:*** Learning readouts of predictive information

***Authors:***

Audrey J Sederberg[1,2,5]

Jason N MacLean[2,4]

Stephanie E Palmer[1,3,4]

1Department of Organismal Biology and Anatomy, University of Chicago, Chicago, IL 60637

2Department of Neurobiology, University of Chicago, Chicago, IL 60637

3Department of Physics, University of Chicago, Chicago, IL 60637

4Committee on Computational Neuroscience, University of Chicago, Chicago, IL 60637

5(current address)Department of Biomedical Engineering, Georgia Institute of Technology, Atlanta, GA, 30332

***Corresponding author***:

Stephanie E Palmer

Address 1027 E 57th St, Anatomy 104, Chicago, IL 60637

Office phone (773) 702-0771

**Abstract.**

To compensate for sensory processing delays, the visual system must make predictions to ensure timely and appropriate behaviors. Recent work has found predictive information about the stimulus in neural populations early in vision processing, starting in the retina. However, to utilize this information, cells downstream must in turn be able to read out the predictive information from the spiking activity of retinal ganglion cells. Here we investigate whether a downstream cell could learn efficient encoding of predictive information in its inputs in the absence of other instructive signals, from the correlations in the inputs themselves. We simulate learning driven by spiking activity recorded in salamander retina. We model a downstream cell as a binary neuron receiving a small group of weighted inputs and quantify the predictive information between activity in the binary neuron and future input. Input weights change according to spike timing-dependent learning rules during a training period. We characterize the readouts learned under spike timing-dependent learning rules, finding that although the fixed points of learning dynamics are not associated with absolute optimal readouts, they convey nearly all the information conveyed by the optimal readout. Moreover, we find that learned perceptrons transmit position and velocity information of a moving bar stimulus nearly as efficiently as optimal perceptrons. We conclude that predictive information is, in principle, readable from the perspective of downstream neurons in the absence of other inputs, and consequently suggests that bottom-up prediction may play an important role in sensory processing.

**Significance statement**

To produce appropriate behavioral responses, such as catch fast-moving prey, the visual system copes with significant sensory processing delays. Spiking activity in the retina captures some of the most predictive aspects of the visual information, but this information must be accessible to downstream circuits. We tested how efficiently predictive information could be read out in downstream neurons and how difficult it is to learn to read out this information, using biologically plausible rules applied only to local inputs. Very simple learning rules could find near-optimal readouts of predictive information without any external teaching signal. Both learned and optimal readouts become less efficient at larger group sizes, suggesting a limit to how many inputs a single output can efficiently read out.

**Introduction.**

To respond efficiently to changing sensory inputs, the brain must predict the future state of the world from past sensory information. For instance, in the salamander visual system, at the minimum such predictions need to compensate for the 50-80 ms processing time of the retina (1) as well as the time for a motor response to be generated. Making these predictions requires leveraging the spatiotemporal structure of the natural world, a computation that is performed efficiently at the first stage of visual processing, in populations of retinal ganglion cells (RGCs) (2). Neurons downstream of the retina likely infer predictions about object motion, but to do so, these downstream cells must learn to read out predictive information from retinal inputs.

The retina has been used as a model to evaluate the theory of predictive coding (3–6), in which deviations from an expected signal are encoded to maximize information transmission efficiency (7–10).  Predictive information (11–13) in the retina, in contrast, is a bottom-up encoding of the predictive aspects of spatiotemporal structure in sensory stimuli (2,14). Depending on context and timescales, it could be advantageous for neural circuits to use the predictability of stimuli in different ways (6,15–23). In the retina, internal temporal correlation in population activity over tens to hundreds of milliseconds can be leveraged to make predictions about the future state of the external world (2). Thus, for early visual processing in particular, the efficient computation of predictive information may be a principal function of retinal circuits and their downstream readouts.

Whether predictive information can be read out and learned by downstream circuits has been comparatively unexplored.  Generally, predictive information is encoded by complex, multi-neuronal activity patterns. To build the range of predictions necessary to bridge the delay between past sensory inputs and future behavioral outputs, downstream circuits must read out predictive information. If the encoding of predictive information is relatively simple, one possible readout mechanism is the perceptron (24): a linear weighting of inputs, followed by a threshold. Such a readout has the advantage of being a biologically feasible, single-step computation, and previous work has shown that perceptrons can efficiently read out predictive information in small sets of cells (2). Here we determine whether optimal readouts of predictive information are learnable and accessible to downstream neurons in the absence of other potentially instructive signals. Because spiking activity of RGCs is the only source of visual information to the brain, we examine whether efficient readouts of stimulus predictive information can be found without direct reference to the stimulus, using instead the internal correlations of RGC population spiking activity.

Finding the prediction-optimal binary readout of population activity in small sets of cells is not trivial because a downstream readout mechanism only has the stream of spikes from RGCs to use as an instructive signal on which to train. More fundamentally, what matters to the survival of the organism is not the optimal readout of spiking activity from a (particular) stimulus for which stimulus information can be explicitly calculated, but instead the optimal readout of predictive information from stimuli it encounters in the natural environment. Supposing this readout function had access to information calculations to determine readout efficiency, the number of possible readout functions for sets of 4 cells exceeds 32,000 (i.e. $2^{2^4-1}$). For sets of five cells, there are more than four billion possible readout functions. This sampling problem is simplified by restricting to readouts to perceptrons, but even then the number of possible readouts is in the thousands for modestly sized cell groups. It would be infeasible for the brain to implement a direct search of this set of possible readouts.

A possible solution lies in taking advantage of the internal predictive information of RGC population spiking activity, which can be thought of as a generalized temporal correlation of RGC population activity. We find that each readout's internal predictive information, the information it has about the future input activity, is also related to that readout's stimulus predictive information, and that optimal readouts of one are likely to be near-optimal readouts of the other. This remarkable fact allows for the efficient encoding of predictive stimulus information by efficient readout of input correlations. Next, we generate optimal readouts of internal predictive information by allowing the linear readout weights to evolve under spike timing-dependent learning rules driven by spiking activity recorded in salamander RGCs stimulated with a natural movie. The specific pattern of readout weights learned depends somewhat on initial conditions, but across sampled sets of cells with a range of initial conditions, we find learned readouts are nearly as efficient as the optimal single-bit readout functions at preserving internal predictive information. Finally, we examine the efficiency of these learned readouts with respect to how well they convey external, stimulus-predictive information. As the input pool grows larger, the learned readouts capture a decreasing fraction of the total input information. However, the stimulus predictive information of the learned readouts approaches that of the optimal readouts, showing that the decrease in efficiency at large group sizes reflects a limit to the compressibility of predictive information. While other schema for reading out predictive information from larger groups of cells will likely be necessary, our results show how local information from small groups of inputs can be used to build efficient downstream readouts of predictive sensory information. These results are most relevant to early visual processing stages beyond the retina.

***Results.***

To investigate whether efficient readouts of predictive information during a natural movie stimulus can be learned, we first established that internal predictive information is a useful proxy for stimulus predictive information. We analyzed optimal readouts of predictive information in a population of larval tiger salamander retinal ganglion cells (n = 53 RGCs) previously recorded using a multielectrode array (2). The retina was stimulated using a moving bar stimulus as well as a video approximating the natural habitat of the larval tiger salamander. Evoked spike patterns were expressed as a binary word across a set of cells, with a 0 for silence and a 1 for spiking activity (1 or more spikes) in each 16.7-ms time bin. Throughout the paper, we will discuss stimulus information (quantified only for activity recorded during the moving-bar stimulus) and internal information (quantified for both stimulus types).

*Reading out stimulus predictive information in a sensory population.*

We compared stimulus information and internal predictive information for spiking activity driven by a moving bar with position and velocity determined by second-order damped harmonic motion driven by random velocity perturbations. That is, the noise-driven dynamics consisted of random velocity "kicks", drawn from a Gaussian distribution at each time step (Fig. 1A, bottom). The bar trajectory was also subjected to deterministic dynamics (phase plot, Fig. 1A, top). Because of the deterministic component of the dynamics, random kicks can trigger a long excursion from the origin (0,0), and during these trajectories, position and velocity of the bar were highly informative of the future bar position and velocity. Over the course of the experiment, the distribution of position and velocity was Gaussian, with average position and velocity equal to 0 (Fig. 1A, shaded contours; scaled coordinates). This represented a prior distribution of stimulus features. The stimulus information of an RGC's spiking activity was determined by how the spike-triggered distribution of position and velocity differed from the prior distribution. While we did not directly estimate stimulus position and velocity from RGC spiking activity here, others have done so with high accuracy (25) using linear decoding (26).

As an example, in Fig. 1B we show four cells, all with sensitivity to particular position and velocity features occurring 100 ms in the past (Fig. 1B), consistent with the 50-100 ms delays incurred by retinal circuitry (1). Spikes in two of the RGCs (RGC 13 and RGC 16) indicated, with high probability, that the bar passed through the center of the image with high speed at t = -100 ms relative to the spike, whereas spikes in RGC 27 and 49 indicated the bar had low velocity and position far from the center (Fig. 1B). All four of these cells were highly informative of past features of the moving bar stimulus: this is visually apparent in the difference between the spike-triggered position-velocity distributions (blue, Fig. 1B) and the prior distribution (gray contours, Fig. 1B). Two of the RGCs (RGC 27 and 49) were also predictive of future stimulus features (Fig. 1B). This can be interpreted in the phase plot of the bar dynamics (Fig 1A). RGC 27 and 49 spiked when the bar was in a region of phase space where the deterministic forces on the bar were strong and the deterministic trajectory traveled far from the fixed point at (0,0), so trajectories were predictable for longer periods of time.

The mutual information of spiking activity and the bar stimulus features (position and velocity) (Fig. 1C) is, graphically, the difference between the spike-triggered distribution (blue) and the prior distribution (gray contours, reproduced from Fig. 1A). Figure 1C quantifies our observations about stimulus information from the spike-triggered distributions in Fig. 1B. While each of these cells were highly informative about past stimulus features, only RGC 27 and RGC 49 conveyed much predictive stimulus information, defined here and throughout the rest of the paper as the mutual information $I(\boldsymbol{x}_t; \{p_{t+dt}, v_{t+dt}\})$ of spiking activity and stimulus position/velocity at relative time $dt$ ( = 1 bin, or + 16.7 ms).

Activity across these four cells was highly informative of future stimulus, but the full four-bit representation of the four-cell activity pattern is highly redundant. By "reading out" only the most predictive spike patterns, nearly all of the predictive information of a cell group can be compressed to a single bit (2). For a set of four cells, there are 16 possible patterns of activity, and if we wish to separate the predictive and non-predictive patterns, there would be $2^{16}$ ways to do so. While the full set of readouts is computationally tractable to sample for N = 4 cells (Fig. S1), it is not for N > 4 cells. A simplification of the problem is to restrict readouts to the class of readouts that can be achieved by perceptrons: a linear weighting of input patterns that is compared with a threshold (Fig. 1D). Two examples of such readouts are explicitly diagrammed for this set of four RGCs (Fig. 1D). Each of these readouts is highly informative of past stimulus features (Fig 1E), but only one readout (green) carried predictive information (Fig. 1F). More generally, a large majority (mean: 78%, SE: 16 % across N = 240 sets) of linear readouts will have stimulus predictive information less than 1 bit/second (Fig. 1H). For this single example with known stimulus feature selectivity, a readout with predictive information of 5 bits/second was found by pooling the most predictive individual cells with similar predictive feature selectivity. However, downstream circuits in the brain must find an effective readout without direct access to stimulus information.

Spiking activity is both predictive of future stimulus features as well as predictive of future spiking activity, and this fact can be used to find efficient readouts of stimulus information. More specifically, information between present and future spiking activity, or internal predictive information, is $I(\boldsymbol{x}_t; \boldsymbol{x}_{t+dt})$, where the "word" $\boldsymbol{x}_t$ is the binary pattern of spikes (1) and silence (0) at time $t$. Across sets of four cells, the stimulus predictive information of a binary word is highly correlated with the internal predictive information (Fig. 1G, r = 0.65, N = 240 sets), meaning that the more predictive spiking activity is of the future stimulus, the more predictive it is of future spiking activity. Moreover, readouts that were highly predictive of the future stimulus tended to have higher internal predictive information (Fig 1H, single set). Across data sets recorded during the moving-bar movie, correlation between internal and stimulus predictive information of linear readouts is high (linear correlation: 0.71 +/- 0.32 (mean +/- SE across sets); N=215 of 240 sets with significant correlation (p < 0.01)). Thus, by finding effective readouts of internal predictive information, the strong relationship between internal and external prediction enabled us to read out stimulus predictive information without having direct access to the stimulus.

This relationship between internal and external prediction was established for data recorded during the moving-bar stimulus. To determine whether this relationship was more general and extended to spiking activity generated by different types of visual stimuli, we evaluated the internal predictive information of spiking activity driven by a natural movie stimulus, a clip of a swimming fish at the 10cm viewing distance of a typical salamander eye (102 repetitions of a 19.2-s clip of the fish movie (Fig. 2A)). Throughout the rest of the paper, a small fish or moving bar icon will be included as an indicator of which stimulus set was used for information calculations. Internal predictive information on the fish movie was highly correlated with the internal predictive information generated by the moving-bar stimulus (Fig 2B; r= 0.82 for sets of 4 cells; = 0.87 for sets of 10 cells; p < 1e-8), and so the correlation between the stimulus predictive information and the internal fish movie predictive information was also high (Fig. 2C; r= 0.65 for sets of 4 cells; = 0.80 for sets of 10 cells; p < 1e-7). Thus, cell sets with high word-word internal information during the natural movie also had high word-stimulus predictive information. However, to find efficient readouts of external predictive information for a particular set of cells using internal predictive information, we must test whether the particular readout compression preserves this useful correlation. Therefore, we compared the predictive information of readouts of internal predictive information during the natural movie to the stimulus predictive information of the same readouts. Similar to the readouts of stimulus predictive information during the moving-bar movie, most readouts of internal predictive information had less than 1 bit/second predictive information (72% +/- 16% SE of all readouts across sets), but a small fraction were highly informative (Fig. 2D, single set example). For this particular set, the predictive information of readouts of internal information were significantly correlated with readouts of external information (Fig. 1E; r = 0.33, p < 1e-4). Across all sampled sets of 4 cells, the linear correlation between internal predictive information and stimulus predictive information was even higher than it was for the sample set (Fig. 2E, sets of 4: mean correlation: 0.73, SE: 0.24; 230 of 240 sets with p < 0.01). Typical correlations between internal and external predictive information readouts were lower but still highly significant for sets of 10 cells (mean correlation: 0.47, SE: 0.30; 231 of 244 sets of 10 cells with p < 0.01).

In summary, neural responses to a moving bar stimulus could predict the future of that stimulus, and the readout of activity of groups of cells that is highly predictive of stimulus features is also highly predictive of the group's future activity, whether that is calculated for a simple moving-bar or a complex natural movie. The stimulus predictive information could be efficiently read out, as could the internal predictive information. An efficient readout under one set of stimulus conditions (natural movie responses) was likely to be a good readout under other conditions (moving bar responses). However, of all possible readouts, only a small minority were efficient readouts of predictive information. We therefore turned to the question of whether efficient readouts of predictive information could be learned using simple, biologically plausible learning rules based on spike timing-dependent plasticity.

*Learning the optimal readouts of internal predictive information*

Predictive information can be thought of as a generalized correlation between the word at time $t$ and at time $t + \Delta t$. By finding efficient readouts with high predictive information, we were identifying the words at time $t$ that were most reliably predictive of a word at time $t + \Delta t$. In other contexts, temporal asymmetry in learning dynamics has been linked to prediction of neural sequences in populations (27). Here specifically, if word A were reliably predictive of word B, then a learning rule that potentiated when word A occurred and depressed synapses when word B occurred could select a readout weight pattern that identified word A and not word B and therefore captured much of the predictive information. We reasoned that an STDP learning rule that relied on relative timing of pre-synaptic and post-synaptic activity could potentially accomplish this, and therefore we simulated learning in which potentiation occurred if an input spike coincides with an output spike and depression occurred if an output spike preceded an input spike (Methods). For each set of cells, we generated a random set of initial patterns of input weights and ran a simulation driven by the spiking data recorded during the natural movie stimulus. The natural movie clip (19.2 s) was repeated 102 times in the experiment, and we drew the training set from half of these clips (Methods). Predictive information was computed on the left-out movie clips. Depending on initial connectivity, one of several final configurations of readout weights was learned, so each set has multiple learned readouts (Fig. 3A; Fig. S1). We quantified efficiency of learned readouts using a firing-rate adjusted metric which compared the predictive information of learned readouts to the highest predictive information of any readout at or below the firing rate of the learned readout (Fig. 3A). This was done to normalize across readouts and to ameliorate biases resulting from readouts that produce more output spikes.

For small sets (4 cells), learned readouts conveyed near-optimal predictive information (Fig. 3B). The average percent of the optimal predictive information learned was 86% (14% SE across N = 240 sets). Firing rates of learned readouts were not saturated at the maximum firing rate and were distributed across the range of readout firing rates, with an average firing rate (mean: 2.3 Hz, SE: 0.9 Hz, N = 240 sets) that was 68% of the maximum firing rate. We quantified the similarity of learned readout rules to the optimal readout rule (Fig. 3C) based on how frequently the rules produced the same output for a given input, weighted by the frequency of the input. Learned readout rules were similar, but not identical, to optimal readout rules (Fig. 3D; mean: 0.71 (SE: 0.20), N = 240 sets). Although they did not precisely match the optimal readout rules, readouts learned under the pair rule were efficient at representing predictive information.

*Learning efficient readouts of up to 10 cells*

Estimating the statistics of anatomical convergence from RGCs, via thalamus, to cortical neurons is difficult knowing only the convergence rates from retina to LGN (estimated at 10-30, (28–30)) and LGN to cortex (estimated at 30, (31)), because convergent thalamic inputs may share more convergent retinal inputs than non-convergent ones. Even without knowing precisely what the biologically relevant RGC pools size are, it is useful to know if efficient readouts can be found for sets larger than four cells. We therefore simulated learning under the pair rule for sets of 7 and 10 cells. For groups larger than 4 cells, assessing learned readout efficiency is more complicated. For groups of 4 cells, it was possible to compute predictive information for all possible readouts, but fully sampling the space of readouts was not possible for larger sets. Based on full samples of sets of four cells, we know that some perceptrons will have near-optimal readout of predictive information, but that there are also some regions of readout space inaccessible to perceptron readouts. We expect that this extends to sets of larger cells, and we limit our comparisons to a sampled subset of perceptron readouts (Fig. S1), with the caveat that we may not have found the optimal perceptrons for larger sets.

Compared to the optimal sampled readouts, we observed a small decrease in readout efficiency (Fig. 4B), from 87% on average for groups of 4, to 82% (SE over groups, 10%, N = 244) for groups of 7, and 80% (SE over groups, 8%, N = 244) for groups of 10. Learned readouts for these cells were less similar to the optimal readout than the readouts for groups of four were (0.63, 0.62 for groups of 7 and 10, respectively; Fig. 4A). While there is a relationship between structural similarity to the optimal rule and efficiency of the readout, many readouts have a high degree of predictive information efficiency with low structural similarity to the optimal rule (Fig. 4C-D; Fig. S2). Thus, for sets of 4-10 cells, efficient readouts of predictive information could be learned without finding the exact structure of the optimal readout.

*Stimulus information of learned readouts*

For sets of 4 to 10 cells, learned readouts captured most of the optimal readout predictive information, measured in terms of the internal predictive information quantified during natural movie stimuli. How efficient are learned readouts at representing stimulus information? To address this, we compare the stimulus information of the learned readouts to that of the full set of cells and to the optimal readouts. For each set of cells, we identified the learned and corresponding optimal internal-information readouts in our simulations (Fig. 5A) and computed for each of these readouts the information about stimulus position and velocity (Fig. 5B). The learned readout stimulus information was compared to the total word-stimulus information (solid line, Fig. 5B) and the optimal readout-stimulus information (dashed blue, Fig. 5B). Because the firing rates of learned and corresponding optimal readouts may change depending on the stimulus type, we compared information efficiency in bits/spike by normalizing by the respective firing rates. We note that the results were qualitatively unchanged without this normalization. Relative to the full cell set stimulus information, learned readouts were much more efficient for groups of 4 cells than for groups of 10 cells. Taken across sets and sampled initial conditions, the median internal-information learned readout of a set of 4 cells was 67% (4%, SEM) as efficient for stimulus information as the full cell set. For readouts of the sets of 10 cells, the efficiency was only 32% (2%, SEM) of the full set (Fig. 5D, solid). However, compared to the optimal readouts, typical learned readouts remained relatively efficient, with an average of 91% (8%, SEM) for the sets of 4 cells and 71% (8%, SEM) for the sets of 10 cells. Moreover, a highly efficient (>95% efficiency relative to the optimal readout) readout pattern was learned for at least one of the simulated initial conditions in the vast majority of sampled sets of cells (223/240 sets of 4; 231/244 sets of 7 and 225/244 sets of 10). In summary, readouts learned under simple spike timing-based learning rules were efficient compared to the best possible single-bit readouts.

## Discussion

Producing successful behavior in an ever-changing environment using sensory information, acquired in the recent past, necessitates prediction at least at one stage of neural processing. Such predictions are enabled by long-range spatiotemporal correlations present in natural stimuli. For example, in recordings of populations of RGCs of salamander retina driven by a simple stimulus with partially predictable dynamics, joint activity patterns transmit information that is predictive of future stimuli (2). However, in order for the organism to make use of this information to predict future sensory stimuli, downstream networks need to be able to be able to learn to read it out. In early sensory processing stages, this learning likely occurs without independent instructive signals providing stimulus information.

We examined whether the temporal correlations within predictive populations of retinal ganglion cells can be used to guide the search for readouts of predictive information. Sets of cells were drawn agnostic to the cell types of constituent cells, though retinal ganglion cell populations are made up of diverse anatomical cell types (32,33) and whether anatomical or functional cell types of cells in a set explain the variability in predictive information across sets and its efficient readout is an interesting question for future work. We showed that groups of cells with high stimulus predictive information also had high internal predictive information, and that while there is not a perfect correlation between the two, efficient readouts of one type of information were likely to be efficient readouts of the other. Potentially, this relationship arises because firing in the retina is maintained by short-term plasticity mechanisms and other network nonlinearities when stimuli are predictable, and is, itself, predictable beyond the observed short correlation time of individual RGCs.

Next, we simulated a set of biologically plausible learning rules based on spike timing dependent plasticity using as “training data” the spiking activity recorded from a set of RGCs driven by a natural movie. This training condition is challenging, but intended to approach a realistic scenario: a downstream neuron reading out bottom-up predictive information does not have independent access to the stimulus or to an error signal to guide its selection of readout weights. While the learning dynamics did not lead to optimal readout weight structure for all initial conditions, the readouts that were found had high information efficiency: relative to the optimal readout of internal predictive information, learned readouts conveyed >80% of the predictive information available to perceptron readouts. Thus, a very simple learning rule, STDP based on pairs of pre- and post-synaptic spikes, found synaptic weights that effectively read out most of the available predictive information.

Finally, we extended our simulations to larger groups of cells. We found that while the efficiency of the readouts relative to optimal single-bit readouts remained high, the efficiency relative to the full cell set decreased for larger cell sets, such that readouts of groups of 10 cells typically preserved 30% of the total predictive information. This suggests a limit to the compressibility of predictive information, and thus an estimate of how many inputs a downstream cell can efficiently read out. In other words, reading out from sets of 4 cells can be accomplished with high absolute efficiency, relative to the full cell set, and perhaps the best way to combine more than four cells is the break this down into indivisible units of four cells, which are later recombined in subsequent processing. Another possibility is that the low efficiency of both learned and optimal readouts at higher $N$ reflects a sampling problem. It could be that restricting our readout mechanism to a simple, perceptron readout is overly limiting, and if we knew how to fully sample the $2^{2^N}$ readouts for a group of $N$ cells, we could read out much more of the predictive information. This is not necessarily the picture painted by the exhaustive sampling of readouts of sets of four cells, but without a guiding theoretical principle of where the optimal bound on readouts of predictive information lies, it is not possible to say with certainty anything about the astronomically under-sampled readouts for sets of 7 or 10 cells.

We emphasized convergence: reading out a single bit from pools of inputs, which would happen somewhere downstream from the retina. This particular dataset was taken from larval salamander, and in the visual system of amphibians, retinal projections terminate in the optic tectum and thalamus (34). Classic work in the optic tectum of amphibians showed that feeding behavior can be evoked directly from electrical stimulation of parts of the optic tectum (35), and perhaps utilizing predictive information is primarily useful for making fast, sub-conscious predictions of the future of sensory stimuli and thus limited to these kinds of automatic, reflex-like behaviors. However, salamander retinal ganglion cells are not unique in encoding predictive information: predictive information is also encoded by rat RGCs (36). In the mammalian visual system, there are both convergence and divergence as visual information passes from the limited number of channels of the optic nerve, to LGN and cortex. This combination may be required to chain together combinations of single-bit readouts that are predictive over larger spatial and temporal regions. In future work, it will be interesting to see if it is possible to build such predictions out of an ensemble of single-bit readouts of many small groups of cells.

## Materials and Methods.

### Multielectrode recordings during movie stimulation of dissected retina.

The dataset used in this study was previously published (2), and complete experimental details can be found in (37). Briefly, a multi-electrode array (252 electrodes, 30 µm spacing) was used to record from a larval tiger salamander retina as images were projected onto the photoreceptor layer. Voltages from the electrodes were recorded at 10 kHz over the course of the multi-hour experiment, and spikes were sorted (37) yielding 53 simultaneously recorded single units.

The movies, referred to as either the naturalistic movie or moving bar, were presented using a 360 by 600-pixel display at 60 frames per second with 8 bits of grayscale. The naturalistic movie was a 19-s clip of fish swimming in a tank with plants in the background and was repeated 102 times. The moving bar was an 11-pixel-wide black bar against a gray background, with dynamics for its position ($x_t^b$, Table 1) and velocity ($v_t^b$) following the equations for a stochastic damped harmonic oscillator:

$$x_{t+\Delta t}^b = x_t^b + v_t^b \Delta t$$

$$v_{t+\Delta t}^{b} = [1 - \Gamma\Delta t]v_t^b - \omega^2 x_t^b \Delta t + \xi_t\sqrt{D\Delta t}$$

The parameters are $\Gamma = 20\text{s}^{-1}$ and $\omega = 2\pi\times(1.5\text{s}^{-1})$, resulting in dynamics that are slightly over-damped: without the stochastic kicks $\xi_t$, bar position would decay back to the center position. The time step $\Delta\text{t} = 1/60\text{s}$, matching the frame rate of the movie. The parameter $D = 2.7\times10^6\text{pixel}^2/\text{s}^3$ was set so that bar position ranges across the screen extent.

The naturalistic movie responses were used for the training dataset and for predictive information calculations (Fig. 2-5), except where specifically noted otherwise. The moving bar responses were used for calculating the information about past and future bar position and velocity (Fig 1, 2, 5).

Sorted spikes were binned into time bins of width $\Delta\text{t} = 1/60\text{s}$. Activity of a set of *m* cells at time $t = n\Delta\text{t}$ is described by the *m*-bit binary word $\boldsymbol{x}_t$. The readout $y_t$ of this set of cells is a binary function on the set of $2^m$ possible binary words. In the case of a perceptron readout (24), this function may be written

$$y_t = \begin{cases} 0 \text{ if } \boldsymbol{w}\cdot\boldsymbol{x}_t \le b \\ 1 \quad \text{otherwise} \end{cases}$$

where $\boldsymbol{w}$ is the length-*m* synaptic weight vector and $b = 1$. We require weights to be excitatory ($w_i \ge 0$).

Throughout the paper, several distinct information theoretic quantities are computed. Word-word internal predictive information is the information the binary word $\boldsymbol{x}_t$ at time $t$ provides about the word $\boldsymbol{x}_{t'}$ at time $t' = t + dt$ for some temporal offset *dt* (38–40):

$$I(\boldsymbol{X}_t;\boldsymbol{X}_{t'}) = \sum_x P_X(\boldsymbol{x}_t)P_X(\boldsymbol{x}_{t'}|\boldsymbol{x}_t)\log_2\frac{P_X(\boldsymbol{x}_{t'}|\boldsymbol{x}_t)}{P_X(\boldsymbol{x}_{t'})}$$

Readout predictive information refers to the mutual information of the perceptron activity $y_t$ at time $t$ and the word $\boldsymbol{x}_{t+dt}$ at time $t' = t + dt$: $I(Y_t;\boldsymbol{X}_{t'})$ Word-word information is symmetric with *dt* ($I(\boldsymbol{X}_t;\boldsymbol{X}_{t+dt}) = I(\boldsymbol{X}_t;\boldsymbol{X}_{t-dt})$, but perceptron-word information is not ($I(Y_t;\boldsymbol{X}_{t+dt}) \neq I(Y_t;\boldsymbol{X}_{t-dt})$. The shorthand “predictive information” is the perceptron-word information for $dt = 1/60\text{s}$ (the temporal bin size). Information was computed using *CDMEntropy*, a Bayesian entropy estimator for binary vector data (41).

*Drawing cell sets.* Cell sets were drawn from the population of 53 total recorded cells. All cells are represented in at least one cell set. To estimate the predictive information bound as a function of firing rate for judging the efficiency of the predictive information of learned perceptrons, we sampled 2k-4k perceptrons for each of the cells sets for which learning simulations were carried out (SI).

**Spike-Timing Dependent Plasticity**

The learning rule is a simplified STDP rule (42,43) adapted for binary neurons and depends on the timing of a single pre-synaptic and a single post-synaptic spike:

$$\Delta \boldsymbol{w}_t = \varepsilon(y_t \boldsymbol{x}_t - \alpha_{\text{LTD}} y_{t-1} \boldsymbol{x}_t)$$

This rule generates potentiation of a weight $w_t^{(i)}$ if the input spike triggered an output spike at time $t$, and depression if an output spike preceded an input spike. We use hard bounds on $\boldsymbol{w}$, $0 < w_i < w_{\text{max}}$ and set $\varepsilon = 0.01$. Practically, because firing is sparse, the maximum weight was chosen to be super-threshold (1.1) which ensured non-zero firing rates of learned readouts. Several variations on this learn rule (triplet-spike (44), homeostatic variations on pair- and triplet-spike rules) were tried, without significant improvement over the results shown (SI).

**Quantifying similarity to the optimal rule**

The similarity to the optimal readout is defined based on the fraction of time bins with one or more input spikes for which the learned and optimal rule produced the same output. Each learned readout corresponds to a set of output rules ($L_i = L(\boldsymbol{x}^{(i)})$), representing the readout response (0,1) to each of the 16 possible input words ($\boldsymbol{x}^{(1)} = 0000$; $\boldsymbol{x}^{(1)} = 0001; \ldots . \boldsymbol{x}^{(16)} = 1111$) (Fig. 3D). Learned rules are sometimes identical to the optimal rules (blue, purple), and sometimes not (green). For each learned readout, similarity between the learned rules $L_i$ and the optimal rules $O_i$ is quantified as

$$\frac{1}{1-p(1)} \sum_{i \neq 1} p(i) \delta_{L_i, O_i}$$

where $\delta_{L_i, O_i}$ is 1 if $L_i = O_i$ and otherwise 0 and $p(i)$ is the probability of observing word $i$ ($\boldsymbol{x}^{(i)}$).

**Acknowledgments**

AJS was supported by the Mary-Rita Angelo Fellowship. SEP was supported by the Alfred P. Sloan Foundation. JNM was supported by National Science Foundation CAREER Award No. 095286.

**References.**

1. Segev R, Puchalla J, Berry MJ. Functional organization of ganglion cells in the salamander retina. J Neurophysiol. 2006;95:2277–92.
2. Palmer SE, Marre O, Berry MJ, Bialek W. Predictive information in a sensory population. PNAS. 2015;112(22):6908–13.
3. Srinivasan M V., Laughlin SB, Dubs A. Predictive Coding: A Fresh View of Inhibition in the Retina. Proc R Soc London B Biol Sci. 1982;216(1205).
4. Hosoya T, Baccus SA, Meister M. Dynamic predictive coding by the retina. Nature. 2005;436(7047):71–7.
5. Kastner DB, Baccus SA. Spatial segregation of adaptation and predictive sensitization in retinal ganglion cells. Neuron. Elsevier Inc.; 2013;79(3):541–54.
6. Berry MJ, Schwartz G. The Retina as Embodying Predictions about the Visual World. In: Predictions in the Brain: Using Our Past to Generate a Future. 2011. p. 295.
7. Bastos AM, Usrey WM, Adams RA, Mangun GR, Fries P, Friston KJ. Canonical Microcircuits for Predictive Coding. Neuron. Elsevier Inc.; 2012;76(4):695–711.
8. Rao RPN, Ballard DH. Predictive coding in the visual cortex: a functional interpretation of some extra-classical receptive-field effects. Nat Neurosci. 1999;2(1):79–87.
9. Kilner JM, Friston KJ, Frith CD. Predictive coding: An account of the mirror neuron system. Cogn Process. 2007;8(3):159–66.
10. Deneve S. Bayesian Spiking Neurons I: Inference. Neural Comput. 2008;20(1):91–117.
11. Bialek W, Nemenman I, Tishby N. Predictability , Complexity , and Learning. 2001;13:2409–63.
12. Chechik G, Globerson a, Tishby N, Weiss Y. Information bottleneck for Gaussian variables. J Mach Learn Res. 2005;6(1):165–88.
13. Creutzig F, Globerson A, Tishby N. Past-future information bottleneck in dynamical

systems. Phys Rev E - Stat Nonlinear, Soft Matter Phys. 2009;79(4):1–5.

14. Salisbury JM, Palmer SE. Optimal Prediction in the Retina and Natural Motion Statistics. J Stat Phys. Springer US; 2016;162(5):1309–23.

15. Berry MJ, Brivanlou IH, Jordan TA, Meister M. Anticipation of moving stimuli by the retina. Nature. Nature Publishing Group; 1999 Mar 25;398(6725):334–8.

16. Trenholm S, Schwab DJ, Balasubramanian V, Awatramani GB. Lag normalization in an electrically coupled neural network. Nat Neurosci. 2013;16(2):154–6.

17. Schultz W, Dayan P, Montague PR. A Neural Substrate of Prediction and Reward. Science (80- ). 1997;275(5306):1593–9.

18. Cooper EA, Norcia AM. Predicting Cortical Dark/Bright Asymmetries from Natural Image Statistics and Early Visual Transforms. PLOS Comput Biol. 2015;11(5):e1004268.

19. Leonardo A, Meister M. Nonlinear Dynamics Support a Linear Population Code in a Retinal Target-Tracking Circuit. J Neurosci. 2013;33(43):16971–82.

20. Borghuis BG, Leonardo A. The role of motion extrapolation in amphibian prey capture. J Neurosci. 2015;35(46):15430–41.

21. Wacongne C, Changeux J-P, Dehaene S. A Neuronal Model of Predictive Coding Accounting for the Mismatch Negativity. J Neurosci. 2012;32(11):3665–78.

22. den Ouden HEM, Daunizeau J, Roiser J, Friston KJ, Stephan KE. Striatal Prediction Error Modulates Cortical Coupling. J Neurosci. 2010;30(9):3210–9.

23. Alink A, Schwiedrzik CM, Kohler A, Singer W, Muckli L. Stimulus Predictability Reduces Responses in Primary Visual Cortex. J Neurosci. 2010;30(8):2960–6.

24. Rosenblatt F. The perceptron: A theory of statistical separability in cognitive systems (Project Para). Buffalo, NY, Cornell Aeronaut Lab. [Washington,: United States Department of Commerce, Office of Technical Services]; 1958;1–59.

25. Marre O, Botella-Soler V, Simmons KD, Mora T, Tkačik G, Berry MJ. High Accuracy Decoding of Dynamical Motion from a Large Retinal Population. PLoS Comput Biol. 2015;11(7):1–25.

26. Bialek W, Rieke F, de Ruyter van Steveninck R, Warland D. Reading a neural code. Science (80- ). 1991;252(5014):1854–7.

27. Abbott LF, Blum K. Functional significance of LTP for sequence learning and prediction. Cereb Cortex. 1996;6:406–16.
28. Cleland BG, Dubin MW, Levick WR. Sustained and transient neurones in the cat's retina and lateral geniculate nucleus. J Physiol. 1971;217(2):473–96.
29. Morgan JL, Berger DR, Wetzel AW, Lichtman JW. The Fuzzy Logic of Network Connectivity in Mouse Visual Thalamus. Cell. 2016;165(1):192–206.
30. Hammer S, Monavarfeshani A, Lemon T, Su J, Fox MA. Multiple Retinal Axons Converge onto Relay Cells in the Adult Mouse Thalamus. Cell Rep. The Authors; 2015;12(10):1575–83.
31. Alonso JM, Usrey WM, Reid RC. Rules of connectivity between geniculate cells and simple cells in cat primary visual cortex. J Neurosci. 2001;21(11):4002–15.
32. Masland RH. The Neuronal Organization of the Retina. Neuron. Elsevier Inc.; 2012;76(2):266–80.
33. Gollisch T, Meister M. Eye Smarter than Scientists Believed: Neural Computations in Circuits of the Retina. Neuron. Elsevier Inc.; 2010;65(2):150–64.
34. Ewert J-P, Schwippert WW. Modulation of visual perception and action by forebrain structures and their interactions in amphibians. EXS. 2006;98:99–136.
35. Finkenstädt T, Ewert J-P. Visual pattern discrimination through interactions of neural networks: a combined electrical brain stimulation, brain lesion, and extracellular recording study inSalamandra salamandra. J Comp Physiol ? A. 1983;153(1):99–110.
36. Salisbury JM., Deny S, Marre O, Palmer S. Predictive information in the retina depends on the stimulus statistics. In: Cosyne. 2016.
37. Marre O, Amodei D, Deshmukh N, Sadeghi K, Soo F, Holy TE, et al. Mapping a Complete Neural Population in the Retina. J Neurosci. 2012;32(43):14859–73.
38. Shannon C. A mathematical theory of communication. Bell Syst Tech J. 1948;27:379-423-656.
39. Cover TM, Thomas JA. Elements of Information Theory. Elements of Information Theory. 2005.
40. Rieke F, Warland D, De Ruyter Van Steveninck R, Bialek W. Spikes: Exploring the Neural

Code. Computational Neuroscience. 1997.

41. Archer EW, Park IM, Pillow JW. Bayesian entropy estimation for binary spike train data using parametric prior knowledge. Adv Neural Inf Process Syst. 2013;26:1700–8.

42. Song S, Miller KD, Abbott LF. Competitive Hebbian learning through spike-timing-dependent synaptic plasticity. Nat Neurosci. 2000;3(9):919–26.

43. Abbott LF, Nelson SB. Synaptic plasticity: taming the beast. Nat Neurosci. 2000;3(november):1178–83.

44. Gjorgjieva J, Clopath C, Audet J, Pfister J-P. A triplet spike-timing-dependent plasticity model generalizes the Bienenstock-Cooper-Munro rule to higher-order spatiotemporal correlations. Proc Natl Acad Sci. 2011;108(48):19383–8.

**Figure legends**

**Figure 1.** Spikes in sets of RGCs are informative of both past and future stimulus information.

A: Phase plot of the dynamics of the moving bar stimulus used in one set of experiments. Shaded area shows the distribution of bar position and velocity over the recording. Bottom: Distribution of stochastic “kicks” to velocity.

B: Cross-section of the spike-triggered distribution of position and velocity for each of four RGCs. Left: Distribution shown for $\Delta t = -100$ ms, which is the position and velocity of the bar 100 ms before a spike occurred. Right: Distribution for $\Delta t = +16.7$ ms.

C: Stimulus information for each of the four RGCs in panel B as a function of peri-spike time.

D: Two candidate readouts of this set of four cells. Solid lines indicate strong connections and dashed lines weak ones.

E: Readout-spike-triggered distribution of position and velocity, for stimulus values at $\Delta t = -100$ ms. Both readouts have high values (5.1 and 5.4 bits/s) of past stimulus information.

F: Readout-spike-triggered distribution of position and velocity, for stimulus values at $\Delta t = +16.7$ ms. Only one of the readouts (green) has high predictive stimulus information (4.2 bits/s).

G: Relationship between word-word internal predictive information and word-stimulus predictive information across sets of four cells.

H: For this particular set of four cells, internal predictive information and stimulus predictive information is correlated across all perceptron readouts.

**Figure 2. Internal predictive information can guide stimulus prediction without explicit reference to stimulus parameters.**

A: Responses of four RGCs (same as in Fig. 1) to a 19.2-s clip of a natural movie, with 102 repetitions of the clip shown. A single frame of the movie is shown.

B: For sets of 4 cells and sets of 10 cells, internal (word-word) predictive information is similar under moving-bar (as in Fig. 1) and natural-movie stimulation conditions.

C: For sets of 4 cells and sets of 10 cells, stimulus predictive information is related to the internal predictive information of the natural movie.

D: Distribution of readout predictive information and plot of predictive information vs firing rate for all single-bit (positive-weight) readouts of internal predictive information, normalized by the full word-word predictive information. Most readouts have low predictive information.

E: Relationship between the readout-word predictive information and the readout-stimulus predictive information for this set of cells.

F: Distribution of linear correlation coefficients between the readout-word predictive information and the readout-stimulus predictive information across sets of 4 and sets of 10 cells. A shuffled distribution in which readout identity was shuffled is shown for sets of 4 cells.

**Figure 3. Near-optimal readouts for groups of four cells are learned under spike-timing-dependent plasticity rules.**

A: Readout-word internal predictive information as a function of firing rate. Learned readouts are black or colored, corresponding to Fig. 3C. The optimal perceptron hull is defined as the highest internal predictive information of any readout at or below a given firing rate.

B: Cumulative distribution of the percent of the maximal predictive information of learned readouts.

C: Readout “rules” of each of the three highlighted learned readouts, juxtaposed with their corresponding optimal readout. Each input word either causes a readout spike (white box) or not (black box). Similarity between two readout rules is the fraction of spiking input words producing matching outputs, weighted by how frequently each word is observed.

D: Cumulative distribution of the similarity to the optimal rule of learned readouts. For sets of four, 39% of learned readouts are the optimal readout.

**Figure 4. Learning near-optimal readouts for larger groups of cells.**

A: Cumulative distribution of the efficiency of learned readouts, as the percent of the optimal readout predictive information attained, across all initial conditions and all cell sets.

B: Cumulative distribution of the similarity of learned readouts to the optimal readout, across all initial conditions and all cell sets.

C-D: Similarity to optimal rule is related to the percent of predictive information read out for sets of 7 and 10 cells.

**Figure 5. Stimulus information of learned readouts is near optimal.**

A: Landscape of possible readout predictive information vs. firing rate for a set of ten cells. One of the learned readouts, and the corresponding local optimal readout, are highlighted.

B: Stimulus information for the full cell set and for the learned and optimal readouts (same cell set as in A).

C: Efficiency of readouts of stimulus predictive information for sets of 4, 7 and 10. Efficiency of typical learned readouts is lower than the efficiency of the full set of 10 cells, but over 70% of the optimal readout efficiency.

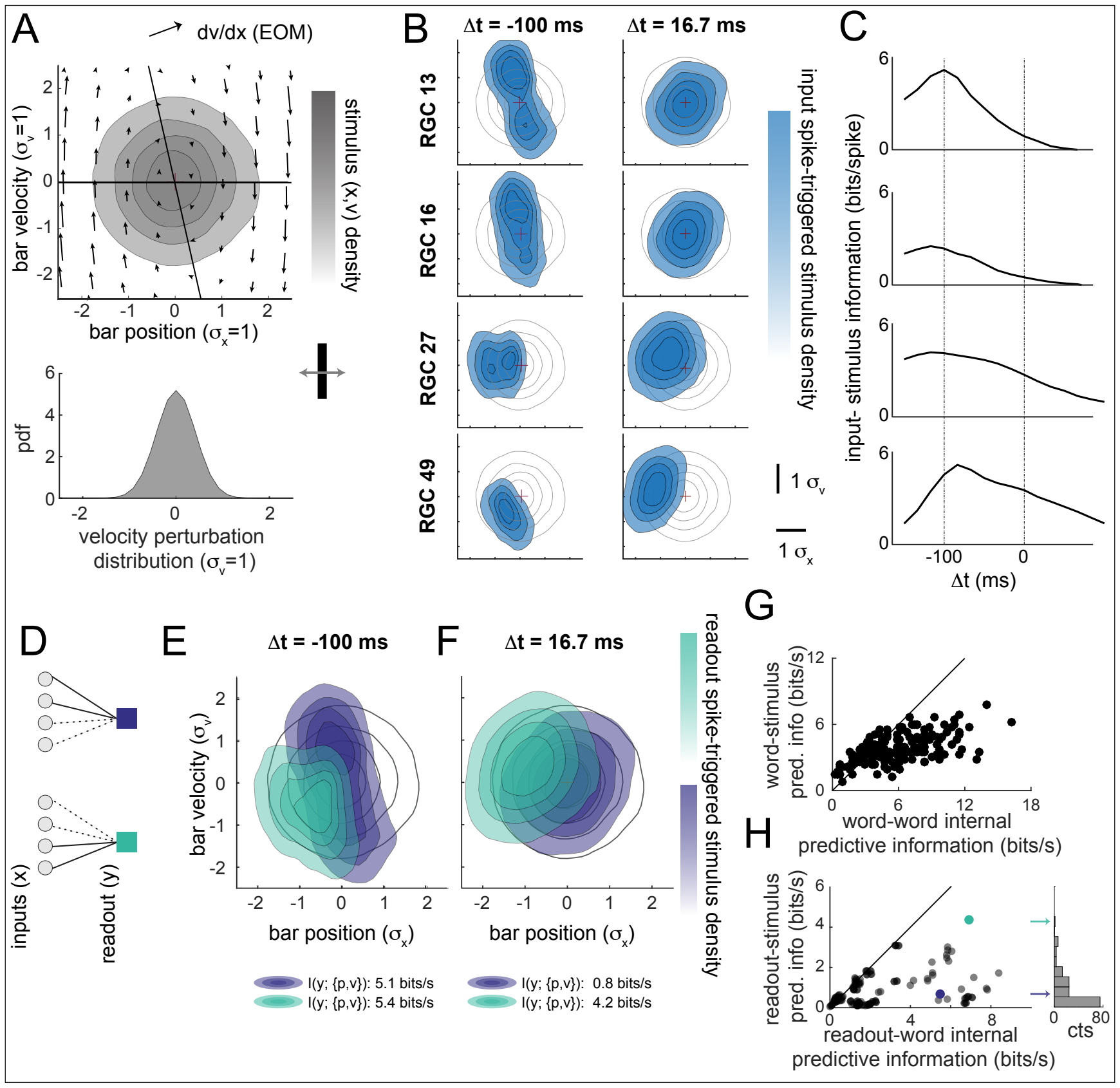

A
dv/dx (EOM)
bar velocity (σv=1)
bar position (σx=1)
stimulus (x,v) density
pdf
velocity perturbation distribution (σv=1)
B
Δt = -100 ms
Δt = 16.7 ms
RGC 13
RGC 16
RGC 27
RGC 49
input spike-triggered stimulus density
1 σv
1 σx
C
input- stimulus information (bits/spike)
Δt (ms)
D
inputs (x)
readout (y)
E
Δt = -100 ms
F
Δt = 16.7 ms
bar velocity (σv)
bar position (σx)
I(y; {p,v}): 5.1 bits/s
I(y; {p,v}): 5.4 bits/s
I(y; {p,v}): 0.8 bits/s
I(y; {p,v}): 4.2 bits/s
readout spike-triggered stimulus density
G
word-stimulus pred. info (bits/s)
word-word internal predictive information (bits/s)
H
readout-stimulus pred. info (bits/s)
readout-word internal predictive information (bits/s)
cts

Figure 1.

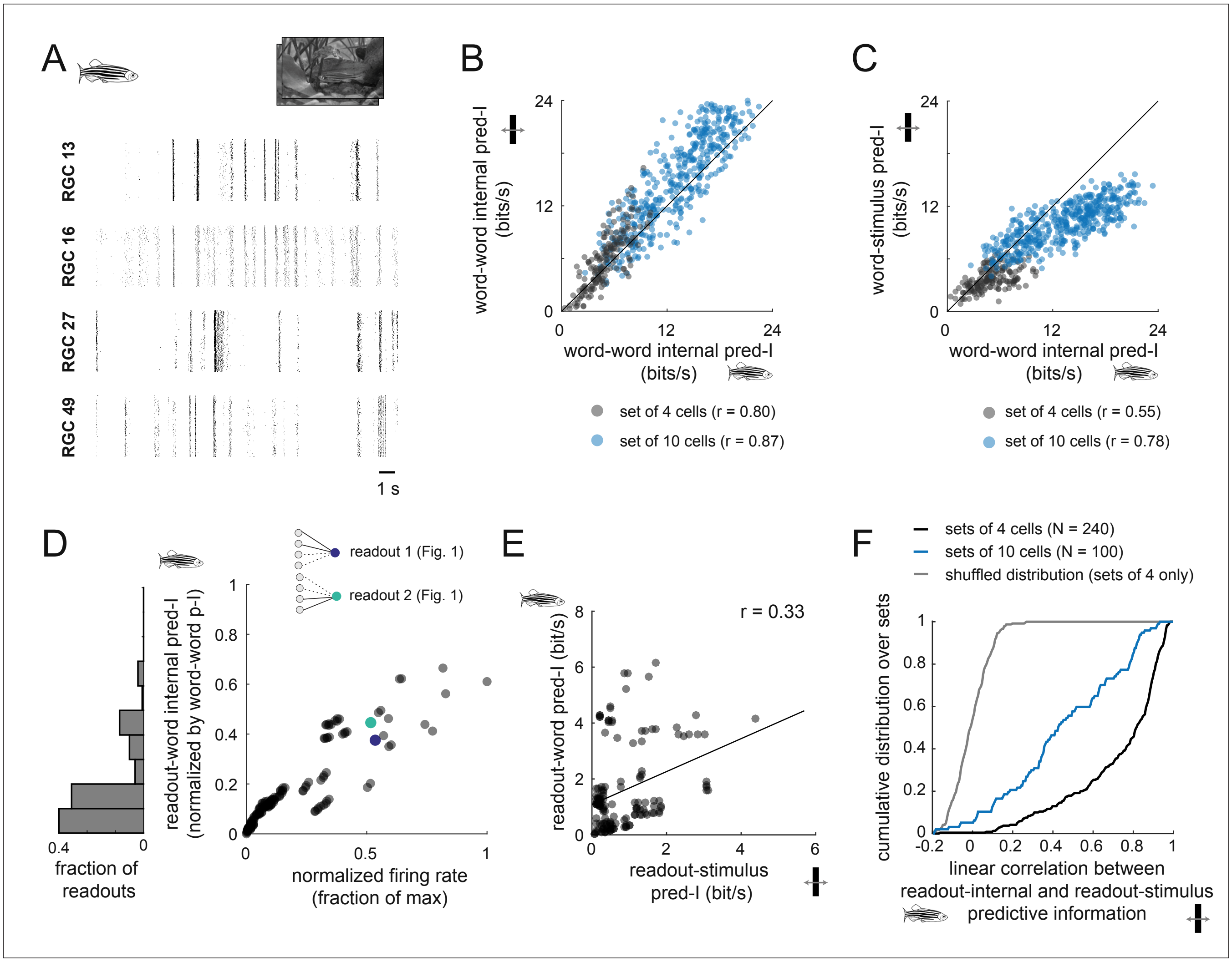
A
RGC 13
RGC 16
RGC 27
RGC 49
1 s
B
word-word internal pred-I (bits/s)
word-word internal pred-I (bits/s)
0
12
24
set of 4 cells (r = 0.80)
set of 10 cells (r = 0.87)
C
word-stimulus pred-I (bits/s)
word-word internal pred-I (bits/s)
set of 4 cells (r = 0.55)
set of 10 cells (r = 0.78)
D
readout 1 (Fig. 1)
readout 2 (Fig. 1)
fraction of readouts
readout-word internal pred-I (normalized by word-word p-I)
normalized firing rate (fraction of max)
E
r = 0.33
readout-word pred-I (bit/s)
readout-stimulus pred-I (bit/s)
F
sets of 4 cells (N = 240)
sets of 10 cells (N = 100)
shuffled distribution (sets of 4 only)
cumulative distribution over sets
linear correlation between readout-internal and readout-stimulus predictive information

Figure 2.

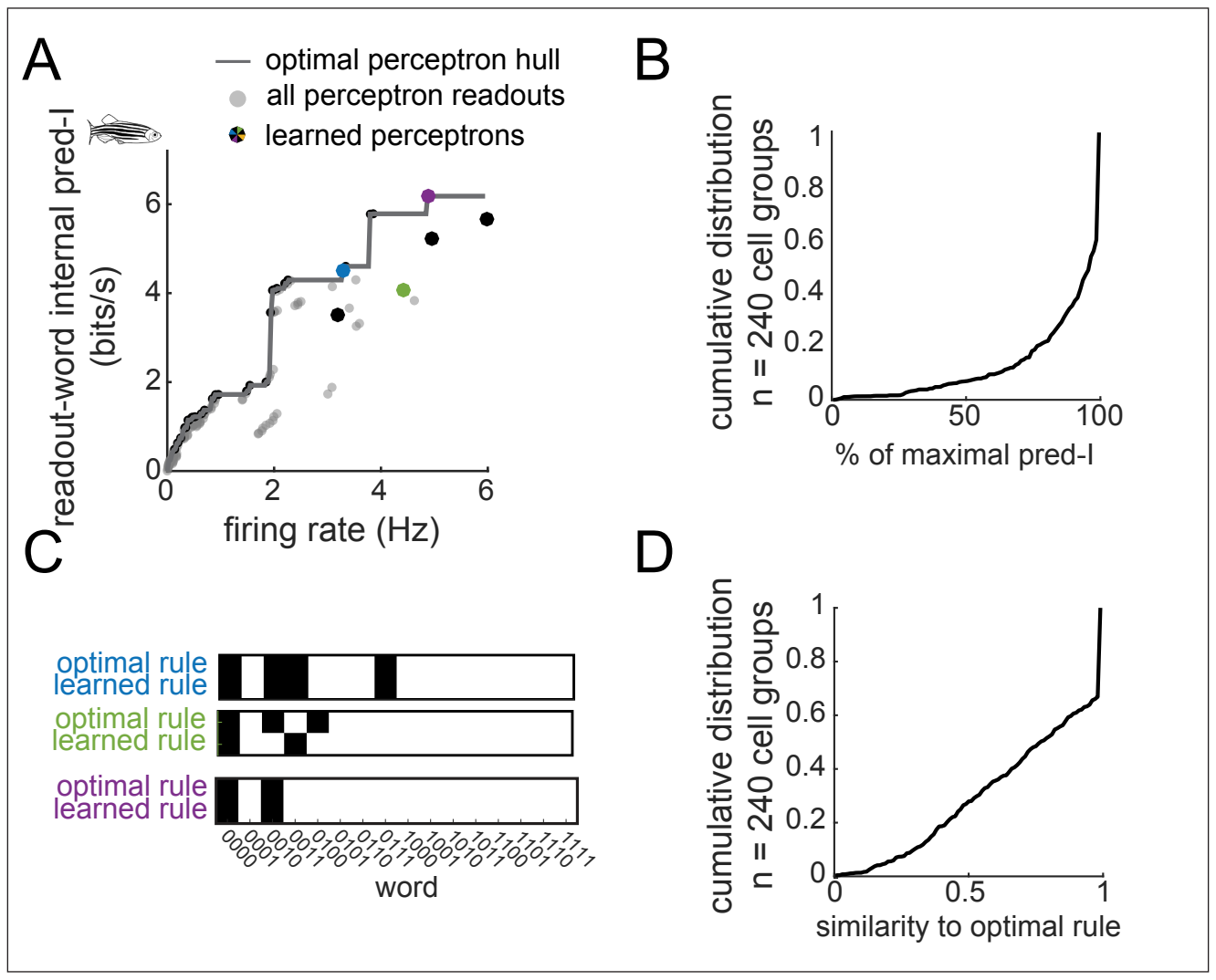
A
optimal perceptron hull
all perceptron readouts
learned perceptrons
readout-word internal pred-I (bits/s)
firing rate (Hz)
B
cumulative distribution n = 240 cell groups
% of maximal pred-I
C
optimal rule
learned rule
optimal rule
learned rule
optimal rule
learned rule
word
D
cumulative distribution n = 240 cell groups
similarity to optimal rule

Figure 3.

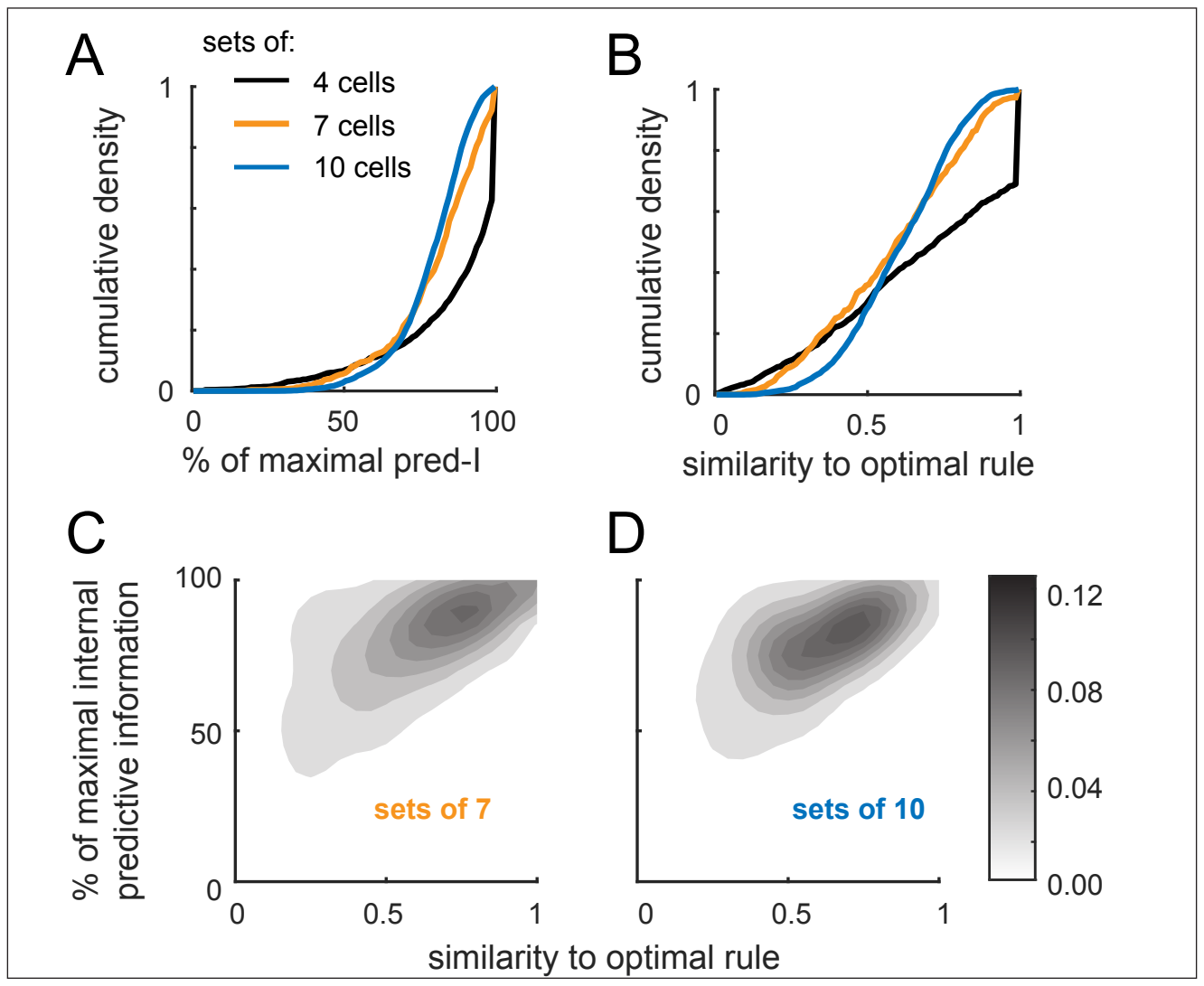

Figure 4.

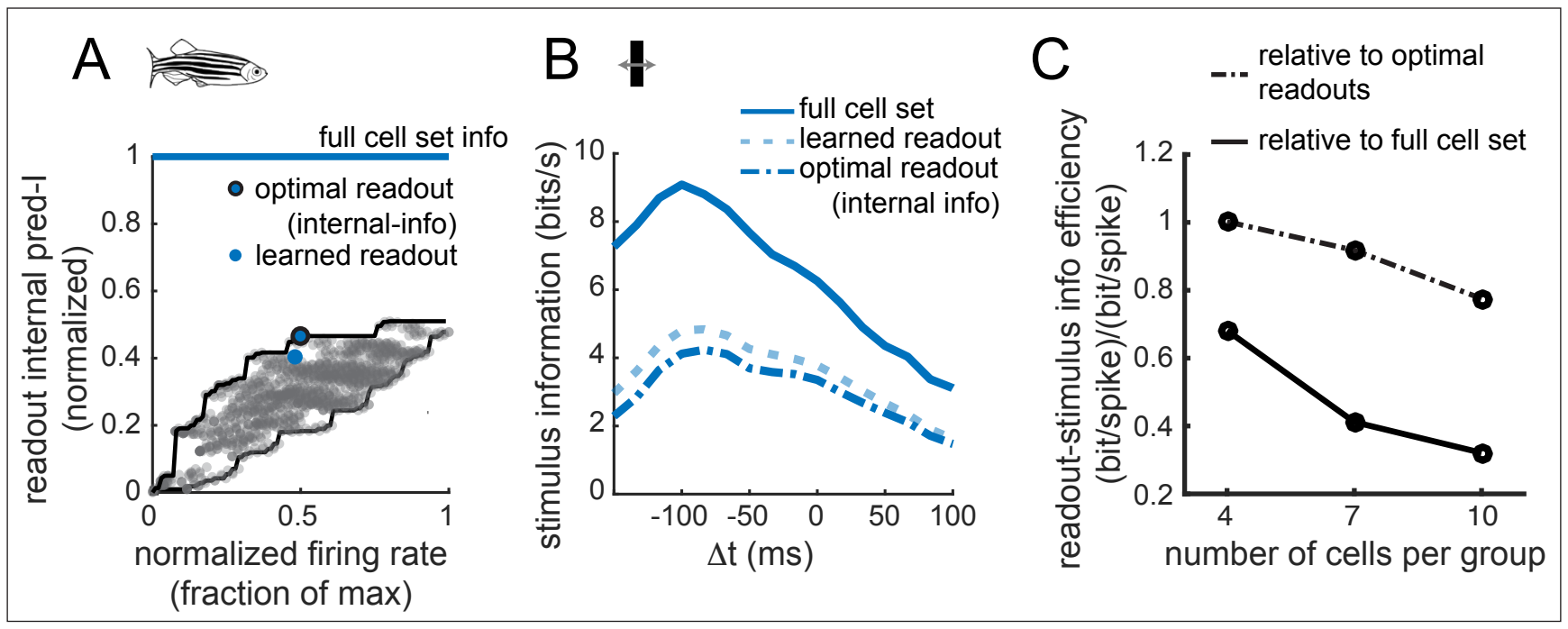

Figure 5.

Supporting Information

*Sampling the space of possible readouts.* (Fig S1)

To judge the efficiency of a learned perceptron for a particular set of cells, we need to compare the learned readout function to the set of possible single-bit readout functions. For a subset of groups of four cells, predictive information was computed exhaustively for each of the 2^15 possible (binary) readout functions (Fig S1A). For these groups of four cells, we computed the maximum readout predictive information as a function of readout firing rate for the exhaustively sampled set of readout functions and for subsets of randomly chosen readouts. At mid-range firing rates (i.e. half of the maximum firing rate), a subsample containing a randomly selected fraction (1%, or 328 of 32768) of all possible readouts produced an estimate of the predictive information bound that was 98% (SE over repeatedly drawn subsample: 3%) of the true bound calculated from the set of exhaustively sampled readouts (Fig S1B).

For sets larger than four cells, it is not possible to exhaustively sample readout functions, and it is possible that a significant gap between the predictive information of the sampled subset of readout functions and the optimal predictive information emerges as the number of cells in the group increases. We partially sampled arbitrary readouts of sets of 5 cells (100,000 samples), including 62 specially chosen partitions in which only a single input word generates an output spike (31 of these) and their converse (31 of these). Over much of the readout firing rate range, such a readout can be found within 10% of the hull generated by sampling 100k partitions. (Fig S1 C-D).

*Learning under other spike timing-dependent rules (Fig. S2)*

We simulated several variations of spike timing-dependent plasticity: pair (see main text) and triplet-spike rules, as well as homeostatic variations of each rule.

The triplet rule depends on the timing of one pre-synaptic and two post-synaptic spikes, such that potentiation is modulated by the post-synaptic inter-spike interval $\Delta_{\text{ISI}}$ :

$$\Delta w_t^{(i)} = \varepsilon\left(y_t x_t^{(i)} \exp\left(-\Delta_{\text{ISI}}/\tau_y\right) - \alpha_{\text{LTD}} y_{t-1} x_t^{(i)}\right)$$

In the triplet rule, potentiation only occurs if there was recently a post-synaptic event at the time of the pre-post pairing. We simulated learning under the triplet rule for $\tau_y = 115$ms and $\tau_y = 167$ms. This rule was defined following Gjorgjieva et al (2011), in which equivalent parameters $\tau_y = 114$ms and $\alpha_{\mathrm{LTD}} = 0.92$. Here $\alpha_{\mathrm{LTD}} = 0.9$, which we found to be where simulation results usually generated weight vectors with both zero and non-zero weights.
For homeostatic learning rules, the sum of weights was constrained to equal 0.75N, where N is the size of the cell group, at each learning timestep.

For each set of cells, we generated a random set of initial conditions, and for each of those initial conditions, four separate learning simulations were carried out, employing each learning rule in turn. We then directly compared the readouts learned under each rule. We visualize this in a matrix, showing for each initial condition of each cell set, which of the four learning rules led to the readout with the highest predictive information (Fig. S2A). Rows are ordered by the predictive information of the full cell set. Most of the matrix is blue, indicating that the pair rule generally led to the highest predictive information in the final readout. For 80% of tested cell sets, the pair rule was optimal for 80% (or more) of initial conditions (Fig. S2B). After the pair rule, the next most common optimal rule was the triplet rule with homeostasis (purple, Fig. S2A, B). We next examined the fraction of recovered predictive information for each simulation under an optimal learning rule. For each cell set, there is a particular pattern of recovered information across all tested initial weight patterns. One difficulty in analyzing these patterns is that the labeling of the four input cells is arbitrary, making comparisons across groups difficult. We noted that the pattern of each set was often most strongly coupled to one of the initial input weights. For each set, we identify the dominant initial input weight based on the correlation of final predictive information with the initial weight strength, and we then order the initial conditions for that set by the initial value of the dominant weight (Fig. S2C). (The same ordering was adopted in Fig. S2A.) The strong patterning in Fig. S2C induced by this ordering shows that having high predictive information in the final learned readout is often dependent on starting with a particular pattern of input weights, specifically with a strong weight in the "dominant" input. However, there is still variability across the cell sets: for many sets, with both small (0.04) and large (0.14) predictive information, nearly all initial conditions reach the same final state. Finally, we examined the distribution of predictive information learned when each rule was optimal (Fig. S2D). Readouts learned under the pair rule have higher median predictive information, but the difference is small: less than 0.01 bits per 16 ms. In summary, the optimal learning rule depends on details of the cell sets and initial conditions, but most often, the simple, pair-spike learning rule produced the most efficient readouts.

*Importance of perceptron structure to perceptron efficiency* (Fig. S3)

Learning the precise structure of the optimal perceptron is only important if information readout depends on finding that optimal structure. Based on the learning results for sets of 4-10 cells, this may not be the case. To see how an efficient readout of predictive information can be found without exactly matching the optimal readout structure, we examined the relationship between the range of readout predictive information and predictive information of the full cell set across many sets of cells (Fig. 5). The range of readout performance (Fig. 5A) was defined as the maximum, over all firing rates, of the difference between the best readout at or below a firing rate and the worst readout at or above a firing rate. This reflected how important finding the optimal structure is: if almost all readouts were within a 20% performance range, then readout structure counted for at most 20% of the readout efficiency. Each set had a particular pattern of predictive information across readouts, but several trends emerged across sampled cell sets. The widest ranges of readout performance were observed when the internal predictive information of the full cell set was smallest (Fig. 5B). For some of these cell sets, the optimal readout captured 90% more information than the worst readout. The range of perceptron performance narrowed to ~20% for word-word information of 0.4 bits/bin (24 bits/s). The overall efficiency of readouts was highest for the lowest internal information, decreasing from 90% for some readouts of internal information of 0.1 bits/bin to 50% for internal information of 0.4 bits/bin (Fig. 5C), which are most often readouts of sets of 10 cells. The difference between the best and worst perceptrons was smaller for these sets, so many perceptrons will tend to have high efficiencies relative to the optimal readout, which is what we saw in the learned perceptrons for sets of 7 and 10 cells: high readout efficiency without high similarity to the optimal rule structure.

*Figure legends.*

SI Fig 1: Sampling the space of possible readouts.

A: Exhaustive sampling of readouts for a single set of 4 cells.

B: Maximum readout information of randomly drawn subsamples vs firing rate relative to the true maximum calculated for the exhaustively sampled set.

C: Random sampling of 100k partitions for a set of 5 cells. Particular readouts with special structure (see text) are highlighted with red circles.

D: Readout information hull calculated for selected readouts relative to the sampled readout hull. 62 of 100000 is massively undersampled, yet these selected readouts are generally closer to the bound than the randomly selected 0.1% of readouts shown in Fig S1B, suggesting a strategy for estimating optimal readout bounds even in the absence of exhaustive sampling.

SI Fig 2: Best rules for learning efficient readouts.

A: The optimal rule at each initial condition for each set of four groups for which learning was simulated, indicated by color. The pair rule without homeostasis (as used in the main text) is most often the best rule.

B: Quantification of A. The pair rule was optimal for 80% of initial conditions in 80% of cell groups. The triplet rule with homeostasis was optimal for 50% of initial conditions in 10% of cell groups.

C: Relationship between initial condition and predictive information of the learned readout.

D: Predictive information of learned readouts was similar across optimal rules but highest for readouts learned under the pair rule.

SI Fig 3: For larger groups, more readouts are near-optimal, but the total readout efficiency decreases.

A: Quantification of the predictive information - firing rate landscape. We characterize each cell group by the maximum fraction of predictive information of any readout, and by the maximum range of readout predictive information at fixed firing rate.

B: The largest range of readout predictive information decreases as total predictive information of the group increases. This means that for cell groups with large total predictive information, most readouts are within 0.2 (as a fraction of the group information) of each other.

C: Larger groups and groups with higher predictive information are less compressible onto a single bit readout of that predictive information. Taken together, for large, predictive groups, the precise structure of the readout weights matters less at a fixed firing rate, so many weights are near optimal. However, the optimal readout gets less efficient.

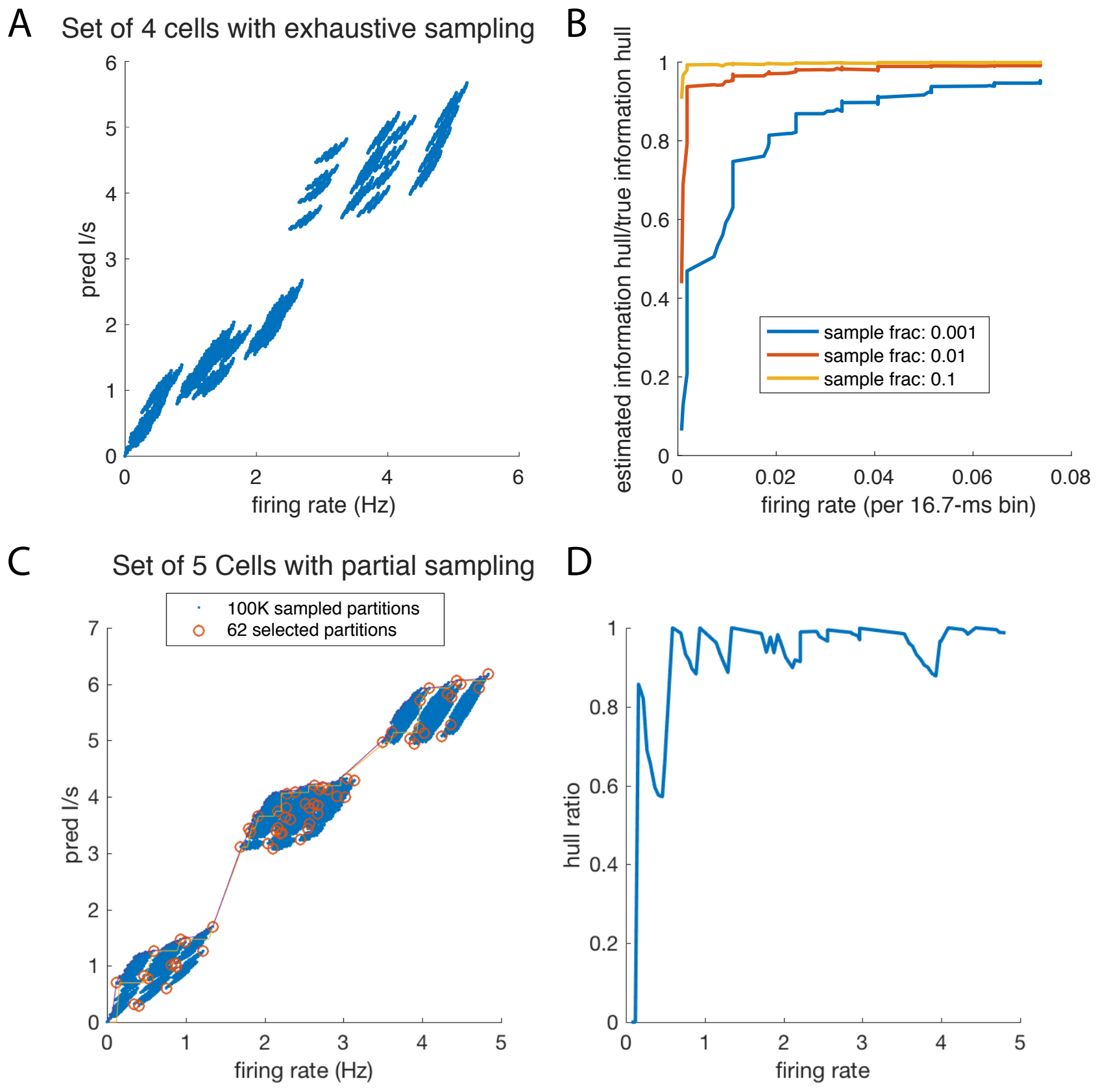

A
Set of 4 cells with exhaustive sampling
pred I/s
firing rate (Hz)
B
estimated information hull/true information hull
sample frac: 0.001
sample frac: 0.01
sample frac: 0.1
firing rate (per 16.7-ms bin)
C
Set of 5 Cells with partial sampling
100K sampled partitions
62 selected partitions
pred I/s
firing rate (Hz)
D
hull ratio
firing rate

Fig S1

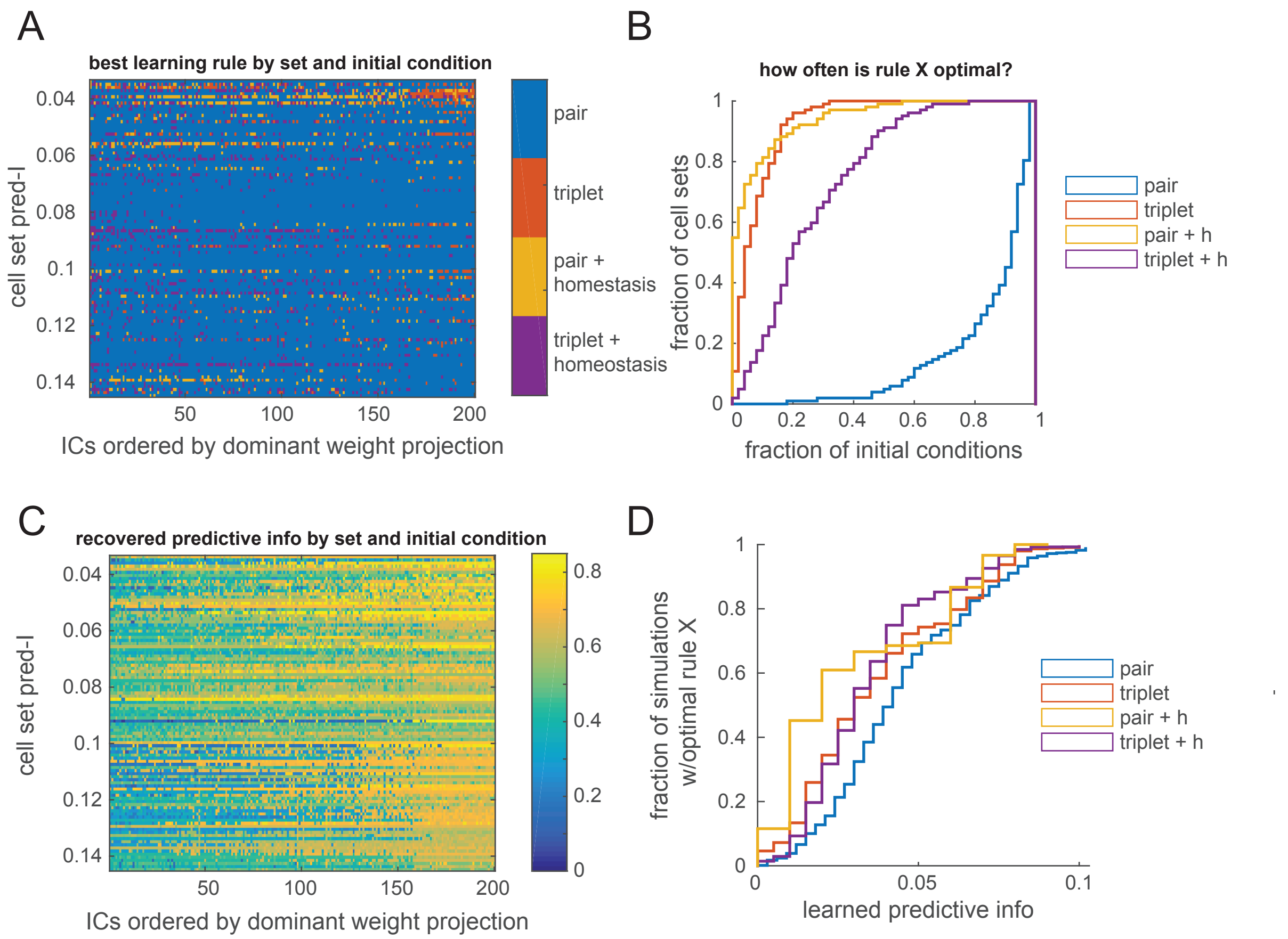
A
best learning rule by set and initial condition
cell set pred-I
0.04
0.06
0.08
0.1
0.12
0.14
50
100
150
200
ICs ordered by dominant weight projection
pair
triplet
pair + homestasis
triplet + homeostasis
B
how often is rule X optimal?
fraction of cell sets
fraction of initial conditions
0
0.2
0.4
0.6
0.8
1
pair
triplet
pair + h
triplet + h
C
recovered predictive info by set and initial condition
cell set pred-I
ICs ordered by dominant weight projection
D
fraction of simulations w/optimal rule X
learned predictive info
0
0.05
0.1
pair
triplet
pair + h
triplet + h

Figure S2.

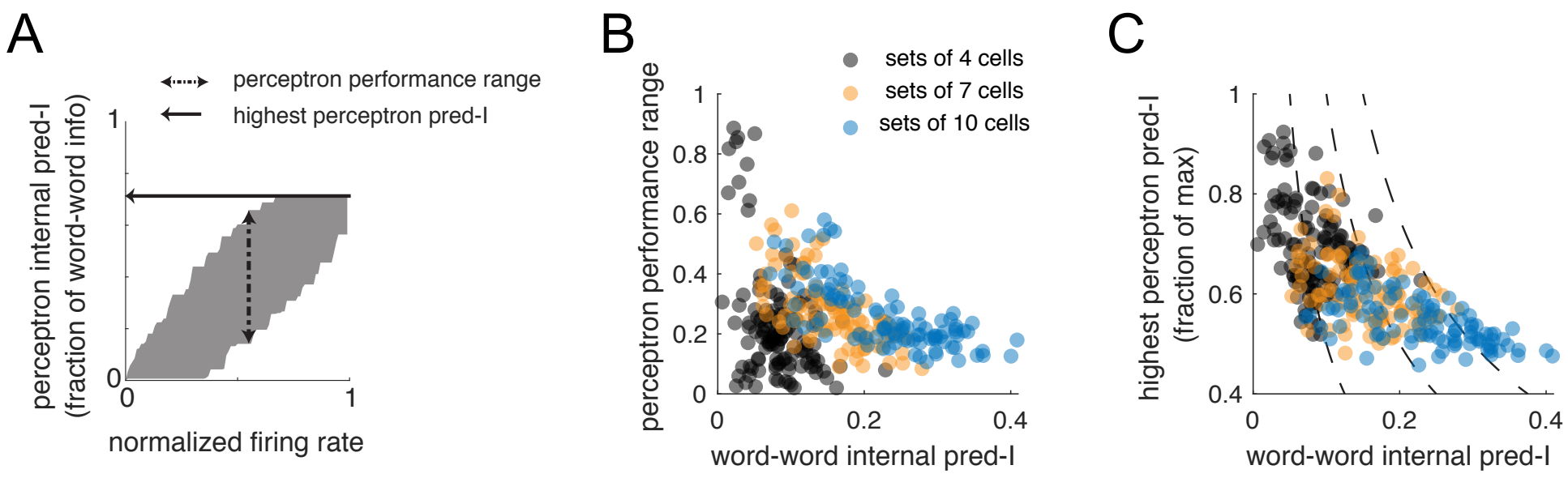
A
perceptron performance range
highest perceptron pred-I
perceptron internal pred-I
(fraction of word-word info)
normalized firing rate
B
sets of 4 cells
sets of 7 cells
sets of 10 cells
perceptron performance range
word-word internal pred-I
C
highest perceptron pred-I
(fraction of max)
word-word internal pred-I

Figure S3.